\documentclass[
reprint,
superscriptaddress,
amsmath,amssymb,
aps,
prl,
twocolumn
]{revtex4-1}

\usepackage{graphicx}
\usepackage{dcolumn}
\usepackage{bm}
\usepackage[breaklinks]{hyperref}
\hypersetup{colorlinks=true, linkcolor=blue, citecolor=blue, filecolor=blue, urlcolor=blue}
\graphicspath{{figures}}
\usepackage{bm}

\usepackage{graphicx,subfigure}
\usepackage{epsfig}
\usepackage{bm}
\usepackage{dcolumn}
\usepackage{color}
\usepackage{physics}
\usepackage{float}
\makeatletter
\newcommand*{\rom}[1]{\expandafter\@slowromancap\romannumeral #1@}
\makeatother
\usepackage{lipsum}
\usepackage{subfigure}
\usepackage{dsfont}
\usepackage{enumitem}
\usepackage{tikz}
\usetikzlibrary{quantikz}

\usepackage[normalem]{ulem}

\newcommand{\CH}[1]{\textcolor[rgb]{1,0.1,1}{#1}}
\begin{document}
\title{Robust Simulations of Many-Body Symmetry-Protected Topological Phase Transitions on a Quantum Processor}

\author{Ruizhe Shen}
\email{e0554228@u.nus.edu}
\affiliation{Department of Physics, National University of Singapore, Singapore 117551}
\author{Tianqi Chen}
\email{tqchen@nus.edu.sg}
\affiliation{Department of Physics, National University of Singapore, Singapore 117551}
\affiliation{School of Physical and Mathematical Sciences, Nanyang Technological University, Singapore 639798, Singapore}
\author{Bo Yang}
\affiliation{School of Physical and Mathematical Sciences, Nanyang Technological University, Singapore 639798, Singapore}
\author{Yin Zhong}
\email{zhongy@lzu.edu.cn}
\affiliation{Key Laboratory of Quantum Theory and Applications of MoE and School of Physical
Science and Technology, Lanzhou University, Lanzhou 730000, People's Republic of China
} 
\affiliation{Lanzhou Center for Theoretical Physics, Key Laboratory of Theoretical Physics of
Gansu Province, Lanzhou University, Lanzhou 730000, People's Republic of China
}
\author{Ching Hua Lee}
\email{phylch@nus.edu.sg}
\affiliation{Department of Physics, National University of Singapore, Singapore 117551}
\date{\today}

\begin{abstract}
Topology and symmetry play critical roles in characterizing quantum phases of matter.  Recent advancements have unveiled symmetry-protected topological (SPT) phases in many-body systems as a unique class of short-range entangled states, notable for their nontrivial edge modes and characteristic ground-state entanglement gap. In this study, we demonstrate the robust simulation of many-body ground states of an Ising-cluster model on a quantum computer. By employing the method of quantum imaginary-time evolution (QITE) combined with enhanced zero-noise extrapolation techniques, we achieve accurate measurements of the transition between trivial and cluster SPT phases. Furthermore, we measured the characteristic edge modes and their associated topological entanglement properties, such as the second R\'enyi entropy, reduced density matrix, and entanglement spectral gap. Our work demonstrates the potential of using QITE in investigating sophisticated quantum phase transitions and critical phenomena on quantum computers.
\end{abstract}

\pacs{}  
\maketitle

\section{Introduction}\label{sec0}
The concepts of topology and symmetry have profoundly guided understanding of quantum phases of matter \cite{hasan2010colloquium,haldane2017nobel,wang2017topological,wen1995topological,levin2006detecting,chen2010local,jiang2012identifying,wen2013topological,wen2017colloquium,wen2019choreographed,lu2022measurement,simon2007pseudopotentials,lee2015geometric}. A particularly compelling focus in recent times is the study of symmetry-protected topological (SPT) phases in many-body systems, particularly those with short-range entangled states marked by robust edge accumulation in the presence of certain symmetries \cite{chen2012symmetry,pollmann2012detection,wen2012symmetry,chen2013symmetry,senthil2015symmetry,verresen2017one,de2019observation,zhang2022digital,zhang2024quantum,tan2025exploring}.  A well-known example of SPT phases is the Haldane phase in spin-1 chains \cite{verresen2017one,scaffidi2017gapless,verresen2018topology}, which exhibits fractionalized spin-1/2
 edge states. Such distinctive properties of SPT phases 
 underscore the critical importance of their experimental investigation and characterization. Symmetry-protected edge or surface states, which are robust against perturbations, offer valuable insights for the development of fault-tolerant quantum computing architectures and advancements in quantum error correction \cite{nautrup2015symmetry,miyake2010quantum}.

In many-body systems, interactions can stabilize intriguing SPT phases, which harbor rich entanglement properties and symmetry-enriched phenomena \cite{parker2019topologically,mesaros2013classification,huang2014detection}.
For a long time, experimentally demonstrating such many-body SPT phases has posed significant challenges. A crucial bottleneck lies in robustly simulating the parent many-body Hamiltonians, which often harbor complicated interactions terms that are elusive in nature. 
In that regard, cold atom systems \cite{see2022many,mukhopadhyay2024observation,qu2013observation,see2024interaction,zhang2008px+,scarola2005quantum,spielman2007mott,mazurenko2017cold,potirniche2017floquet,song2018observation,brown2019bad,sompet2022realizing} such as Rydberg atom simulators \cite{lee2011antiferromagnetic,de2019observation,de2019observation,browaeys2020many,samajdar2021quantum,bluvstein2021controlling,qin2024kinked,myerson2022construction,shen2023proposal} have emerged as powerful platforms in many-body simulation, underscored by their success in demonstrating dynamical SPT phases in various experiments \cite{song2018observation,de2019observation,sompet2022realizing}. However, these platforms face limitations due to challenges in interaction engineering \cite{chin2010feshbach}, which hinder the further characterization of more sophisticated many-body SPT phases.

As cutting-edge quantum simulators, digital quantum computers have recently emerged as promising candidates that can potentially transcend these challenges \cite{preskill2018quantum,kim2023evidence} with their exceptional scalability and precise control over qubit interactions. This makes them ideal tools for simulating quantum many-body systems, as evidenced by recent breakthroughs in realizing nonequilibrium Floquet SPT phases \cite{dumitrescu2022dynamical,zhang2022digital}. Despite notable advancements in digital quantum simulations \cite{abrams1997simulation,fauseweh2024quantum,smith2019simulating,koh2022stabilizing,smith2022crossing,koh2022simulation,koh2024realization,shen2311observation,chen2024direct,chen2023robust,shen2024enhanced,guo2021stark,frey2022realization,zhu2021probing,xu2018emulating,guo2021observation,chen2022error}, the simulation of many-body ground states on quantum computers remains in its nascent stages \cite{chen2022high,ghim2024digital,poulin2009preparing,stanisic2022observing}.  This is primarily due to the lack of efficient quantum algorithms capable of directly obtaining the ground state of a many-body Hamiltonian.  While various approximation methods for accessing ground states have been proposed and implemented \cite{poulin2009preparing,motta2020determining,nam2020ground}, these techniques often struggle to achieve high-fidelity ground states and lack robustness against device noise.

\begin{figure*}
\centering
	\includegraphics[width=0.7\linewidth]{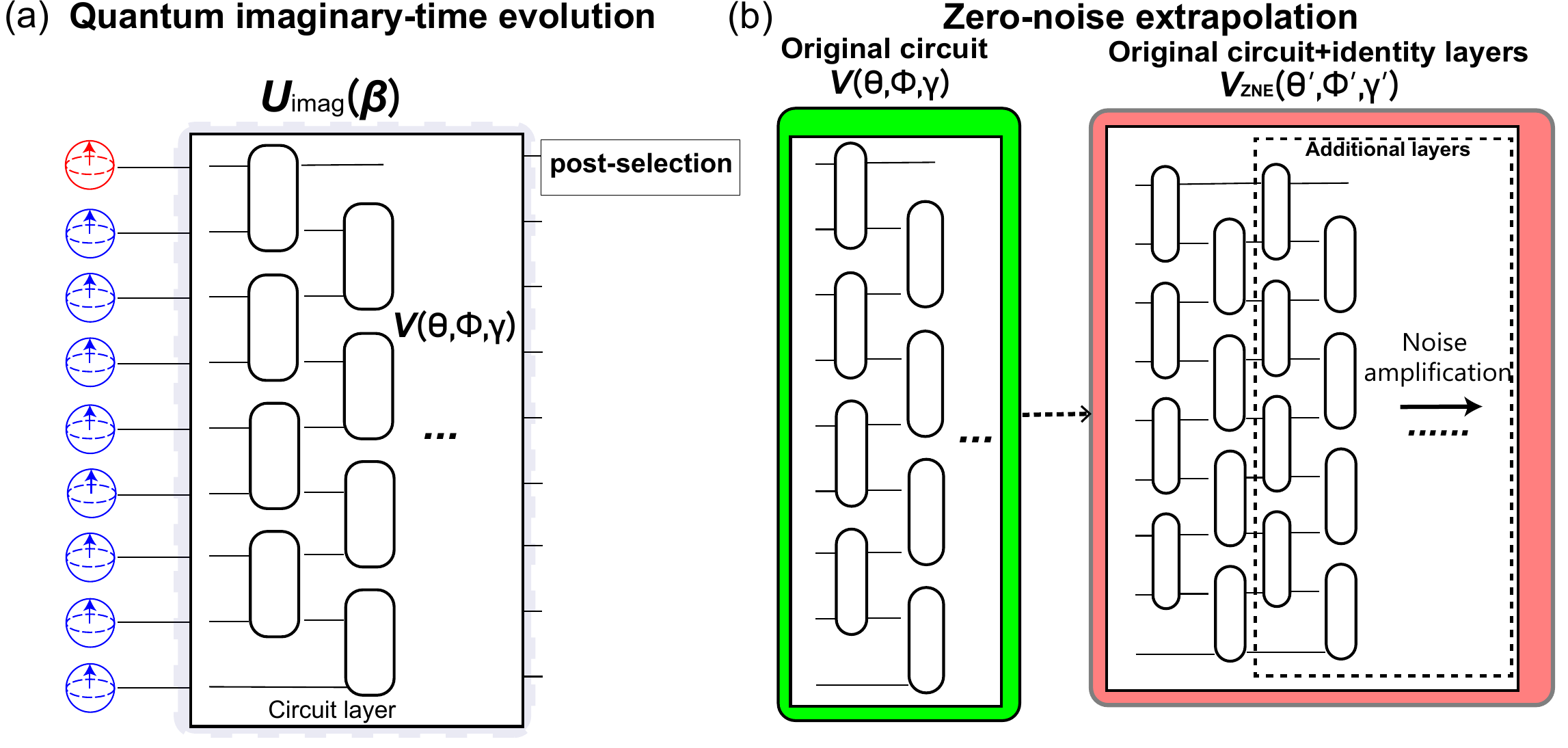}
	\caption{ {\bf Quantum circuits for our quantum imaginary-time evolution (QITE) with zero-noise extrapolation (ZNE).}  In (a), we present the schematic of a circuit designed to implement the nonunitary evolution $e^{-\beta H}$ with imaginary time $\beta$. This setup couples the non-unitary system of interest (depicted in blue) with a single ancilla qubit (depicted in red), such that the overall evolution is unitary. 
Post-selection, represented by the operator $\ket{0_{A}}\bra{0_{A}}$, is applied to the ancilla qubit to recover the non-unitary evolution. The whole process described by $U_{\rm Imag}(\beta)$ in Eq.~\ref{qite2} is implemented through a parameterized circuit with layers arranged as illustrated, with each unit $V(\theta,\phi,\gamma)$ characterized by trainable parameters $\theta,\phi,\gamma$ that are variationally optimized. In (b), we illustrate the structure of our variational-based ZNE approach. The original circuit $V(\theta,\phi,\gamma)$ (green) is extended to form the ZNE circuit $V^{\rm ZNE}(\theta^{\prime},\phi^{\prime},\gamma^{\prime})$ (red), which should produce the same output under noiseless conditions. However, with more variational layers, the $V^{\rm ZNE}(\theta^{\prime},\phi^{\prime},\gamma^{\prime})$ circuit amplifies noise effects. By tuning the number of additional identity layers, results with zero noise can be extrapolated. Further details on the design and application of these circuits are provided in Methods.    }
	\label{circuit}
\end{figure*}

In this work, we demonstrate the robust digital quantum computer simulations of complicated many-body ground states, focusing on the cluster SPT phase in particular.  To efficiently access ground states,  we utilize an enhanced ancilla-based Quantum Imaginary-Time Evolution (QITE) technique \cite{mcardle2019variational,motta2020determining,lin2021real,kamakari2022digital}, combined with tailored variational quantum error mitigation strategies.

A key feature of our approach is the integration of variational simulation with tailored zero-noise extrapolation (ZNE) \cite{he2020zero}. This significantly offsets inevitable noise effects in present-day quantum processors, offering more robust measurements compared to using the variational quantum eigensolver (VQE) approach alone \cite{peruzzo2014variational,kandala2017hardware}. 
This enhancement enables us to accurately simulate transitions from trivial to cluster SPT states and to identify nontrivial edge states on quantum processors. Furthermore, we were able to access the symmetry-protected entanglement properties of the cluster SPT phase. With these advancements, our work paves the way for 
physically exploring a broad range of critical phenomena that exist in complicated many-body models \cite{ashida2017parity,xiao2019observation,liu2021non,qin2025dynamical,shen2022non,yang2024non,liu2024non,xue2024topologically,shen2023proposal}.

\begin{figure}[t!]
	\centering  
	\includegraphics[width=1\linewidth]{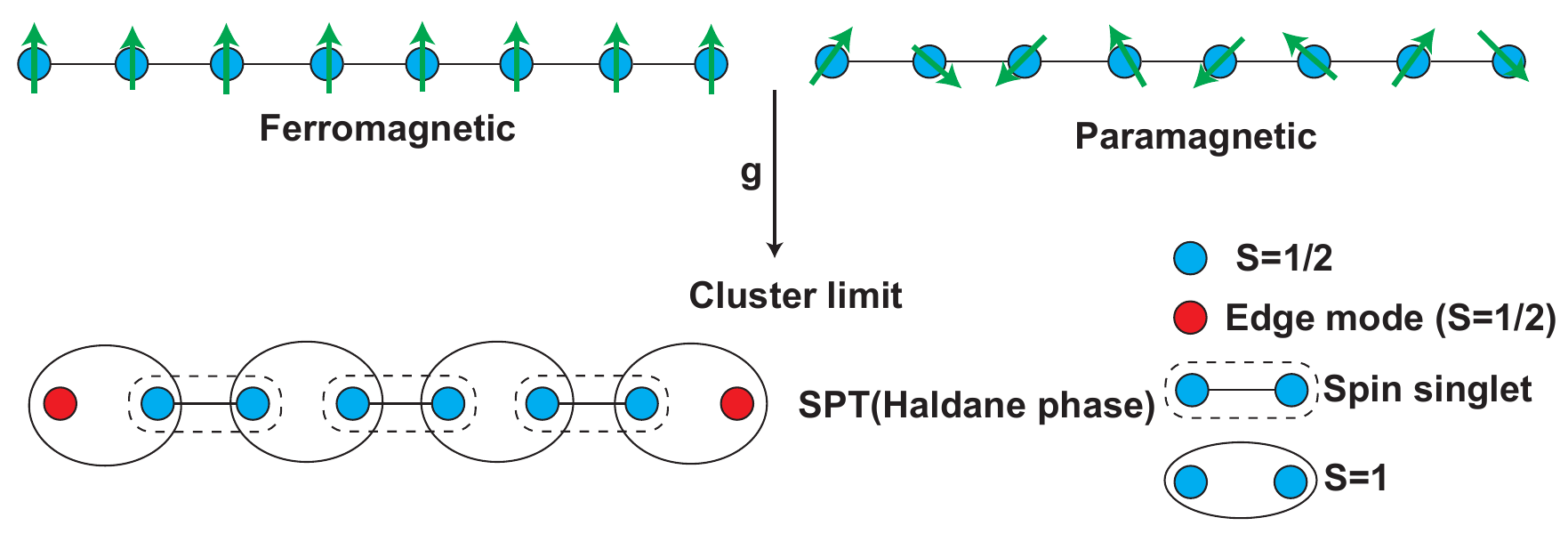}
	\caption{Schematic of the phase transition from trivial (ferromagnetic or paramagnetic) to cluster symmetry-protected topological (SPT) phases in the Ising-cluster model [Eq.~\eqref{spt}], driven by increasing the strength of the three-body cluster interaction $g$. In the cluster limit, the system transitions into the cluster SPT phase, also known as the Haldane phase, characterized by the emergence of edge modes (red node, $S=1/2$). The structure of this cluster SPT phase is as follows: two linked blue $S=1/2$ spins form a spin singlet, and two unlinked $S=1/2$ spins within a solid circle combine to form an $S=1$ spin. The two red $S=1/2$ spins are unpaired and serve as edge modes.
 }
	\label{fig:carton}
\end{figure}

\section{Results}

\subsection{Methods for robust ground-state simulations}\label{sec2}

In this section, we outline a general framework that we employed for simulating robust ground states on a noisy quantum processor. This framework consists of two key components: ground state preparation and its error mitigation.

To prepare the ground state of an arbitrary Hamiltonian $H$ with $L$ qubits, we employ the QITE approach. This technique propagates an initial state  $\ket{\psi}_{0}$ towards the ground state with the lowest eigenenergy through the following nonunitary dynamics $\ket{\psi}_{\bf ground}\approx e^{-\beta H}\ket{\psi}_{0}/\sqrt{||e^{-\beta H}	\ket{\psi}_{0}||}$, where $\beta$ is the imaginary time. With sufficiently large $\beta$, the evolved state converges to the ground state.

\begin{figure*}
	\centering
\includegraphics[width=1\linewidth]{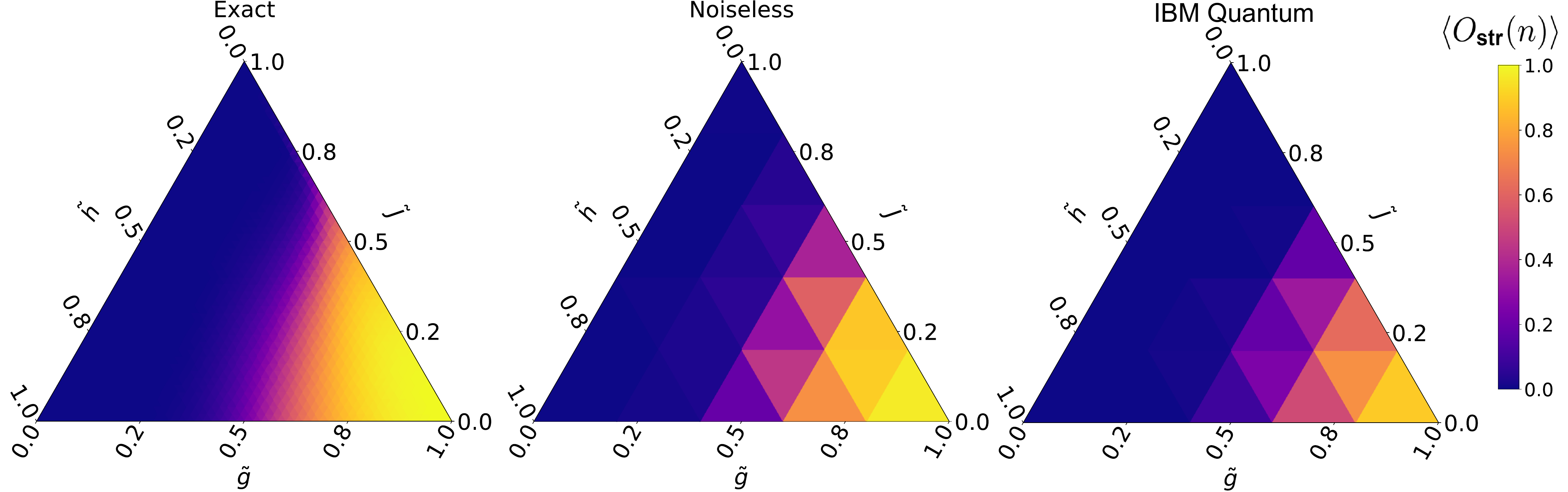}
\caption{{\bf Robust measurement of the Ising-cluster phase diagram.} We present the phase diagram in the normalized parameter space $(\tilde{J},\tilde{h},\tilde{g}) = (J, h, g)/(J + h + g)$. The SPT phase region is determined by measuring the string-order parameter $\left\langle O_{\bf str}(n)\right\rangle$  [Eq.~\ref{str}] with $n=L$, which assumes the non-trivial value of 1 (yellow) in the SPT phase in the lower right corner. Results from the noiseless simulator agree well with exact classical calculations, as expected, and also with measured values from our IBM Quantum simulation despite significant device noise. Evidently, our approach successfully captures the general phase diagram both in principle (noiseless simulation), and in practice (IBM Quantum simulation). For all results, the system size is $L=8$ spins.}
	\label{fig:phasediagram}
\end{figure*}

To implement this process on a quantum circuit, which can only simulate unitary processes, we embed the $L$-qubit nonunitary operation $e^{-\beta H}$ into an $(L+1)$-qubit extended unitary operation $(U_{\rm Imag}^{\dagger}U_{\rm Imag}=I)$, defined as:
\begin{equation}
U_{\rm Imag}(\beta)=ue^{-\beta H}\otimes {\ket{0_A}\bra{0_A}}+U^{\prime},  
\end{equation}
where the $U^{\prime}$ terms restores the unitarity of the overall operation, and $u$ is a scaling factor (see  {Methods}). The desired outcome is obtained by performing post-selection on the ancilla qubit $A$ in the state $\ket{0_{A}}$. This process is described as:
\begin{equation}\label{qite2}
	\ket{\psi}_{\bf outcome}={\rm Post}[(U_{\rm Imag}(\beta )\ket{\psi_{0}  }\ket{0_{A}})],
\end{equation}
where $U_{\rm Imag}(\beta)$ denotes the coupling between the physical chain and ancilla qubit, and the process ``${\rm Post}$'' is for the post-selection on $0_{A}$, as shown in FIG.~\ref{circuit} (a).

In our work, we utilize trained variational circuits to perform the quantum simulations. By organizing circuits into stacked layers of basic single-qubit and entangling gates, we achieve a substantial reduction in circuit depth, greatly enhancing the robustness of our simulations against inevitable device noise. Additionally, our variational framework is integrated with the zero-noise extrapolation (ZNE) technique \cite{he2020zero}, where the number of layers (and hence the amount of noise) is varied to allow extrapolation of the results towards the zero noise limit. This process is illustrated in FIG.~\ref{circuit} (b). Importantly, this approach admits a control knob for the extent of noise, whose effects can be understood better even though they cannot be eliminated.  Detailed descriptions of the implementation of the QITE and ZNE methods are provided in the Methods and Supplementary Information.

\subsection{Model}\label{sec1}
An archetypal SPT phase is the Haldane phase, characterized by a spin-1 bulk with spin-1/2 edge modes. While theoretical research has extensively explored the Haldane-type SPT phase in various dimensions \cite{scaffidi2017gapless}, the architectural constraint of most existing superconducting quantum processors makes the one-dimensional version most practical for simulations. In this study, we focus on the Haldane-type SPT phase on a one-dimensional spin-1/2 chain \cite{scaffidi2017gapless,duque2021topological}, described by the following Ising-cluster Hamiltonian:

\begin{equation}\label{spt}
{H=-J\sum_i  Z_i Z_{i+1}-h\sum_i  X_i-g\sum_i  Z_{i-1} X_i Z_{i+1},}
\end{equation}
where $J$, $h$, and $g$ represent the strengths of the Ising interactions, transverse fields, and cluster interactions, respectively. Here, we use the Pauli operators, defined as $X=\ket{0}\bra{1}+\ket{1}\bra{0}$, $Y=-i\ket{0}\bra{1}+i\ket{1}\bra{0}$, and $Z=\ket{0}\bra{0}-\ket{1}\bra{1}$. 

An intriguing feature of this model is the cluster SPT phase, characterized by a nontrivial global  $\mathbb{Z}_{2}\otimes \mathbb{Z}_{2}$ symmetry.  To understand this global symmetry,  we begin by analyzing this model under the cluster limit at $J=h=0$. In this regime, the Ising-cluster Hamiltonian can be simplified to $H_{c}=\sum_iZ_{i-1} X_i Z_{i+1}$. The global symmetry of this cluster Hamiltonian can be identified using spin-flip operators acting on odd and even sites, defined as $P_{\rm odd}=\prod_{n\in {\rm odd}}X_{n}$ and $P_{\rm even}=\prod_{n\in {\rm even}}X_{n}$. The commutation relations $[P_{odd}, H_{c}]=[P_{even}, H_{c}]=0$ indicate that this system is protected by a $\mathbb{Z}_{2}\otimes \mathbb{Z}_{2}$ symmetry, and $\sum_{i}X_{i}$ fields do not break this symmetry.

We then extend the above scenario to include cases with nonzero Ising interactions. While Ising interactions breaks the $\mathbb{Z}_{2}\otimes \mathbb{Z}_{2}$ symmetry established by $P_{odd}$ and $P_{even}$, the SPT phase remains protected by two other symmetries: the $\mathbb{Z}_{2}$ symmetry, represented by the global spin-flip operator ${P}={P}_{odd}{P}_{even}$, and the time-reversal symmetry, characterized by $T=K$, where $K$ denotes complex conjugation.
Thus, the SPT phase in the Ising-cluster model [Eq.~\ref{spt}] is not restricted to the special case at $J=h=0$. Instead, it supports a $\mathbb{Z}_{2}\otimes \mathbb{Z}_{2}^{T}$ SPT phase across an extended parameter regions \cite{verresen2018topology,son2011quantum,duque2021topological},  {as we later present and map out through quantum device measurements}.

\begin{figure*}
\centering
\includegraphics[width=1\linewidth]{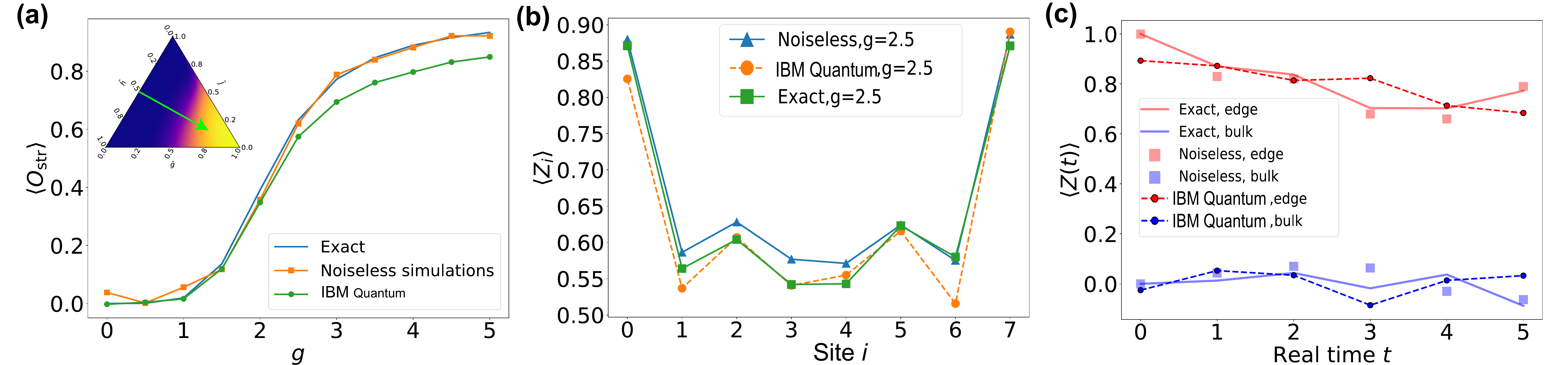}
\caption {{\bf Measurement of robust cluster edge modes.} Panel (a) presents quantum and classical measurements of the string-order parameter  [Eq.~\ref{str}] with $n=L$, with a clear transition between trivial ($\left\langle O_{\bf str}(n)\right\rangle\approx 0$) and non-trivial ($\left\langle O_{\bf str}(n)\right\rangle\approx 1$) SPT phases as $g$ increases, with $J=h=1$ kept constant. This corresponds to the green arrow trajectory in the inset phase diagram (taken from FIG.~\ref{fig:phasediagram}), which is plotted against the normalized parameters $(\tilde{J},\tilde{h},\tilde{g}) = (J, h, g)/(J + h + g)$. 
Panel (b) displays the measured $Z$-magnetization [Eq.\ref{z}] in the symmetry-protected topological (SPT) phase with $g=2.5$, which is prominently larger at the edges, indicative of topological edge localization.
In panel (c), we examine the edge (red) and bulk (blue) magnetization evolution during the quenching process $e^{-itH}\ket{0111110}$, governed by the cluster SPT Hamiltonian with $g=2.5$.  These results reveal that edge modes remain notably stabilized over time.
Other parameters for the results in (b) and (c) are $J=1$ and $h=1$. The length of the Ising-cluster chain is $L=8$ qubits for all results.
}
\label{fig:edge1}
\end{figure*}

Prior Floquet realizations of the Haldane phase \cite{dumitrescu2022dynamical,zhang2022digital}, based on Trotterized dynamics, primarily focus on nonequilibrium behaviors. Here, our study focuses on the phase transitions occurring in equilibrium ground states, which have not been comprehensively explored in digital quantum simulators. This transition is driven by the interplay between cluster interactions and Ising interactions or $X$ fields, which can trigger the phase transition from trivial to $\mathbb{Z}_{2}\otimes \mathbb{Z}_{2}^{T}$ SPT phase, also recognized as one of the Haldane-type SPT phases \cite{verresen2017one,scaffidi2017gapless,verresen2018topology}. In the following text, we refer to this specific SPT phase, protected by $\mathbb{Z}_{2}\otimes \mathbb{Z}_{2}^{T}$ symmetry, as the cluster state.

A hallmark of the transition from trivial to cluster SPT phases is the emergence of edge modes, characterized by the presence of spin excitations near boundaries. This phenomenon is depicted in the diagram presented in FIG.~\ref{fig:carton}, where edge modes of unpaired red spins become notably pronounced at boundaries under significant cluster interactions. This characteristic feature is further demonstrated in subsequent simulations, where spin density profile is directly measured.

\subsection{Measurement of the SPT transition phase diagram}

Following the above techniques and theoretical preparation, we present the result of our quantum simulation on the IBM Quantum processor. Here, we first identify the transition between the cluster SPT and trivial phases. In one-dimensional systems, nonlocal string order parameters can indicate the presence of cluster states. To achieve this, we utilize the following parameter, defined as \cite{smith2022crossing,verresen2021gapless}:
\begin{equation}\label{str}
	O_{\bf str}(n)={Z}_0{Y}_1 [\prod^{n-3}_{k=2}X_{k}]Y_{n-2}Z_{n-1},
\end{equation}
In the cluster limit of $g\rightarrow1$ and $J\rightarrow0$, we expect to have $\langle O_{\bf str}(n) \rangle=1$. Departures from this value hence serve as a robust indicator of the SPT-to-trivial phase transition.

We present both classically computed and quantum computer measured phase diagrams in FIG.~\ref{fig:phasediagram}  \cite{duque2021topological}.  To mitigate finite-size effects, which arise from limitations in {the} circuit layout and gate errors, we extend the range of the string-order operator to its full length, with $n=L$ in Eq.\ref{str}. As shown in FIG.~\ref{fig:phasediagram}, both noiseless simulations and exact theoretical results converge towards the value of $1$ near the cluster limit, confirming the presence of the cluster SPT phase.  

To more precisely characterize the veracity of our phase transition measurement, we choose a path (taken as $h=J=1$ for simplicity) between the trivial and nontrivial cluster SPT phases, and measure the string order parameter [Eq.~\ref{str}] along it, as the interaction strength $g$ is varied. As presented in FIG.\ref{fig:edge1} (a), the transition is evident, with a rapid increase from  $\langle O_{\bf str}(n) \rangle\approx 0$ to  $\langle O_{\bf str}(n) \rangle\approx 1$ as $g$ increases. Due to the finite size, this increase is rapid but not sharp, as also reflected in the exact calculations (blue), that agree almost exactly with our noiseless simulations. Even though the measured IBM quantum results slightly depart from the exact results, it is still able to very accurately replicate the location of the jump. 
 Importantly, despite inevitable noise on physical quantum processors, our noisy simulations effectively and accurately capture the complete phase diagram across distinct phases, with the boundary between the topological and trivial phases clearly observed.

\subsection{Measurement of symmetry-protected edge states}\label{sec4}

In the above section, we present the measurement of the Ising-cluster model's phase diagram through its string order parameter. 
Below, we proceed to measure its SPT-nontrivial edge modes through the $Z$-magnetization of our obtained ground state $\ket{\psi_{g}}$: 
\begin{equation}\label{z}
\langle Z_{i}\rangle=\bra{\psi_{g}}Z_{i}\ket{\psi_{g}}.
\end{equation}
Indeed, edge state accumulation can be observed in the non-trivial parameter region where the string order parameter is significant. Accordingly, we utilize a setup with $g=2.5$ with $J=h=1$, where the string order parameter reaches a  {reasonably} high value. 
In FIG.~\ref{fig:edge1} (b), we present the measured $Z$-magnetization profile of an $8$-qubit physical chain, which vividly illustrates distinctive edge modes. 
Importantly, the noiseless simulation (blue curve), noisy simulation (yellow curve), and exact results (green curve) all demonstrate very good consistency across most qubits, achieved through our improved variational ZNE method (see Supplementary Information).

We further demonstrate the robustness of cluster SPT edge modes by simulating the unitary quenching dynamics $\ket{\psi(t)}=e^{-itH}\ket{0111110}$, starting from an edge-excited state. This robustness is analyzed through the total edge  $\langle Z_{\rm edge}(t)\rangle$ and bulk $\langle Z_{\rm bulk}(t)\rangle$ magnetization:
\begin{equation}\label{edge}
\langle Z_{\rm edge/bulk}(t)\rangle=\sum_{i\in {\rm edge/bulk}}\bra{\psi(t)}Z_{i}\ket{\psi(t)}/N_{\rm edge/bulk},
\end{equation}
where the cluster SPT Hamiltonian is set at $g=2.5$ and $J=h=1$, and $N_{\rm edge/bulk}$ represents the number of edge or bulk spins. As shown in FIG.~\ref{fig:edge1} (c),  despite a slight decay in the edge magnetization (red curves) over the time scale $t\in[0,5]$, edge modes remain remarkably robustly stabilized with $\langle Z_{\rm edge}(t)\rangle>0.6$.
 By contrast, for the bulk sites that initially exhibit vanishing $Z(t)$, their magnetization remains hovering around zero despite the presence of significant quantum device noise.

\begin{figure*}
	\centering
	\includegraphics[width=0.99 \linewidth]{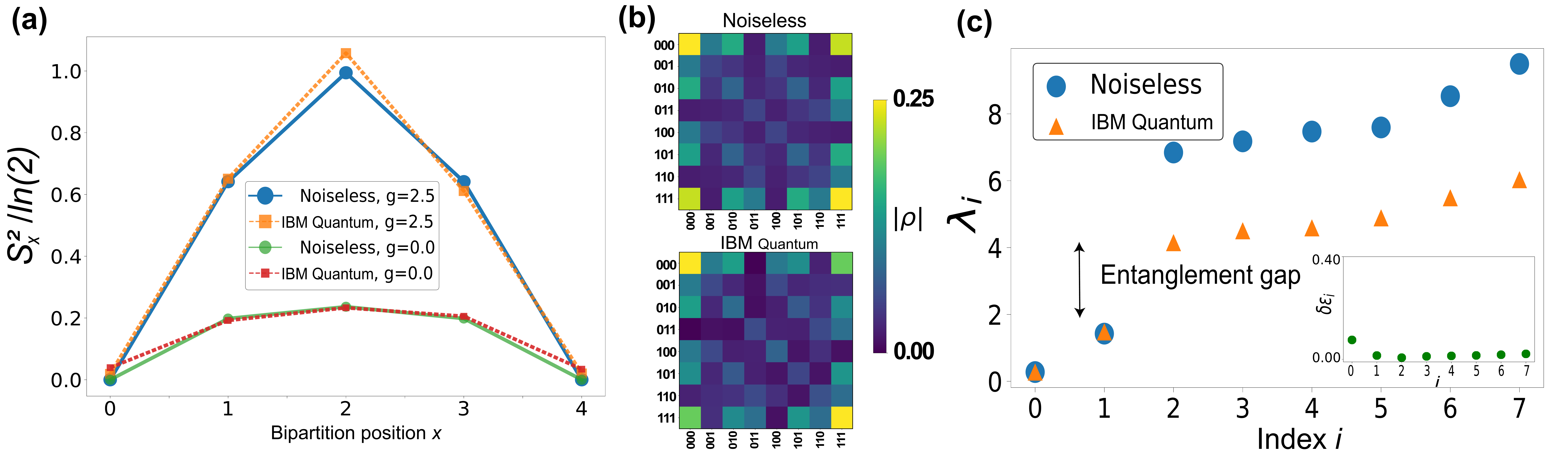}
	\caption{{\bf 
	Measurement of many-body entanglement properties. 
	}
	(a) Measurement of second-order Rényi entropy at various entanglement subsystem sizes $x$ (see further details in Method) in a $L=4$ physical chain, with very good agreement between noiseless simulations and IBM hardware measurements. In the topological phase with symmetry protection ($g=2.5$, the blue and yellow curves),  the half-chain entropy ($x=2$) approaches the ideal value of  $S^{2}_{L/2}= \ln{2}$, 
	much higher than that of the trivial phase at $g=0$.	
	(b), Measurement of the reduced density matrix  {$\rho_A$} at non-trivial $g=2.5$. The  {subregion} $A$ is constructed from the first three qubits of a $L=4$-qubit chain. The entire matrix is derived through the decomposition of Pauli strings [see Eq.~\ref{rho}], also with very good agreement between noisy and noiseless results. 
(c), Entanglement spectrum $\lambda_{i}=-\ln(\epsilon_{i})$. $\epsilon_{i}$ denote the eigenvalue of $\rho$, derived from the data in (b). Most salient is the characteristic entanglement gap between $\lambda_{1}$ and $\lambda_{2}$. The apparent discrepancies in the simulated (noiseless) and hardware (IBM) $\lambda_{i\geq 2}$ actually correspond to very small measurement discrepancies $\delta\epsilon_{i}=|\epsilon^{\rm IBM}_{i}-\epsilon^{\rm Noiseless}_{i}|$ as shown in the inset. This is due to the very small values of $\epsilon_{i\geq 2}$ in the reduced density matrix, which is of the order of the noise contributions in (b). For all results,  $J=h=1$. 
	}
	\label{en}
\end{figure*}

\subsection{Measurement of symmetry-protected topological entanglement}\label{sec6}

In the above simulation, we demonstrated the transition between topological and trivial phases. To further establish our realization of the cluster SPT phase, we measure its entanglement properties, specifically the second Rényi entropy and entanglement gap.

In the cluster SPT phase, nontrivial topological entanglement arises from presence of irremovable mid-gap edge states, and a distinct entanglement gap emerges as one of its hallmarks \cite{qi2012general,pollmann2010entanglement,turner2011topological,dalmonte2018quantum} (also see Methods). In small systems, weak residual correlations between the edge states slightly lift their degeneracy, but the entanglement gap importantly still persists. Furthermore, in the cluster phase, these edge spins remain isolated, leading to characteristic entanglement entropy of $\ln(2)$ \cite{pollmann2010entanglement,azses2020identification}. 

In general, we can measure the second Rényi entropy in a system as follows. Defining a bipartition of the system into subsystems A and B, the reduced density matrix of subsystem A is obtained by tracing out the other subsystem B: $\rho_A=\operatorname{Tr}_B(|\psi\rangle\langle\psi|)$. The second Rényi entropy is then given by $S_{A}^{2}=-\log(\operatorname{Tr}[\rho_{A}^{2}])=-\log(R_{A}^{2}),$
where $R_{A}^{2}=\operatorname{Tr}[\rho_{A}^{2}]$ is the trace of the squared reduced density matrix \cite{horodecki2002method,cornfeld2019measuring,pichler2013thermal,islammeasuring,azses2020identification}.

Directly measuring the elements of a reduced density matrix presents challenges due to high computational or experimental costs, as well as susceptibility to sampling errors (see Supplementary Information for details). To circumvent these challenges, we first adopt an alternative strategy that directly measures the second Rényi entropy instead of the reduced density matrix \cite{johri2017entanglement,xue2024topologically,daley2012measuring}. This method involves preparing a two-copy state $\ket{\psi}=\ket{\psi}_{1}\ket{\psi}_{2}$, where $\ket{\psi}_{1}=\ket{\psi}_{2}$ represents the desired ground state. The trace of a length-$x$ subsystem can then be measured through a subsystem swap operation, defined as:
\begin{equation}\label{r2}
R^{2}_{x}=\bra{\psi}_{1}\bra{\psi}_{2}\operatorname{SWAP_{sub}(x)}\ket{\psi}_{1}\ket{\psi}_{2}
\end{equation}
Here, the operation $\operatorname{SWAP_{sub}(x)}$ swaps a selected length-$x$ subsystem within $\ket{\psi}_{1}$ with the corresponding subsystem in its copy $\ket{\psi}_{2}$ \cite{islammeasuring,azses2020identification}. This approach significantly improves robustness against sampling errors. Additionally, we use the quantum amplitude estimation (QAE) technique to measure $R_{x}^{2}$, which reduces the required number of measurements on the qubits, as detailed in the Methods.

In FIG.~\ref{en} (a), we present the noiseless simulation and measured (IBM Quantum) results for the second Rényi entropy $S^{2}_{{x}}=-\ln{R_{x}^{2}}$ as a function of truncation position $x$, using our direct measurement strategy. The alignment between the noiseless and noisy results highlights the efficacy of our measurement technique. A key feature of the cluster SPT phase is the presence of topological degeneracy, which manifests as two nearly identical eigenvalues in the half-chain reduced density matrix, $\epsilon_{1}\approx\epsilon_{2}=1/2$, where $\epsilon_{i}$ is the $i$-th  {largest} eigenvalue of $\rho_{{A}}$. All other eigenvalues remain relatively insignificant. 

While this degeneracy is not exact in small systems since opposite edge states are not perfectly decoupled due to their finite separation, as long as its protection symmetry remains unbroken, the entanglement gap is preserved. Therefore, we expect the half chain second Rényi entropy to be $S^{2}_{\rm {x=L/2}}\approx\ln(2)$, which we accurately measured as shown in FIG.~\ref{en} (a) \cite{azses2020identification} (see Methods).

However, to measure the entanglement spectrum and obtain the entanglement gap, it is necessary to map out (i.e., perform quantum state tomography for) the reduced density matrix. As discussed, such measurements based on quantum state tomography are susceptible to significant sampling errors \cite{cramer2010efficient,christandl2012reliable}. To mitigate this, we concentrate on a manageable system of four physical sites and consider the subsystem A to be composed of the first three physical qubits. Details of quantum state tomography are provided in the Methods section.  We present the  {noiseless (simulated) and noisy (measured)} results of $|\rho_{\CH A}|$ in FIG.~\ref{en} (b), which exhibit substantial consistency between each other.

Finally, we reveal the symmetry-protected topological entanglement gap by computing the entanglement spectrum, as shown in FIG.~\ref{en} (c). Upon obtaining the matrix elements for $\rho_{A}$, as shown in FIG.~\ref{en} (b), we diagonalize $\rho_{A}$ and obtain the entanglement spectrum $\lambda_{i}=-\ln(\epsilon_{i})$, where $\epsilon_i$ are its eigenvalues arranged in decreasing order. Notably the first two low-lying $\lambda_0,\lambda_1$ exhibit excellent agreement between their noiseless (blue) and measured (orange) results, with a prominent entanglement gap clearly separating these states from the higher levels $\lambda_{i\geq 3}$. Although there appears to be a significant difference between the noiseless and measured results for these $\lambda_i$, in reality the measurement discrepancy $\delta\epsilon_{i}=|\epsilon^{\rm IBM}_{i}-\epsilon^{\rm Noiseless}_{i}|$ (green dots) is very small, as shown in the inset. This is because these $\epsilon_{i}$ are very close to zero and are thus extremely susceptible to small discrepancies in the matrix elements of $\rho_A$ due to sampling noise. Further improvement on this accuracy is possible but would require a substantially larger number of measurements, which is costly  (see Supplementary Information for detailed estimations). 
Despite these limitations, it is apparent that a distinct entanglement gap has been measured, therefore strongly confirming the symmetry-protected topology of the ground state at $g/J=2.5$ (see Methods).

\section{Discussion}\label{dis}
In this work, we achieved a significant advancement in the robust digital simulation of nontrivial phase transitions.  
A key accomplishment of this study is the robust simulation of many-body topological phase transitions realized through the integration of QITE and ZNE methods.  This framework provides enhanced resilience against device noise compared to traditional approaches, such as the use of well-established VQE and quantum annealing approaches without advanced error mitigation strategies. 
Building on this enhanced framework, our results 
encompasses the intricate characterization of phase transitions, the identification of nontrivial edge states, and the measurement of entanglement properties. These underscore the potential of digital quantum simulations for performing sophisticated and robust measurements in complex many-body systems.
Furthermore, as variational methods are widely employed in many-body simulations, our improved ZNE technique offers an efficient approach for achieving high-fidelity results on current noisy quantum hardware \cite{giurgica2020digital,he2020zero,pascuzzi2022computationally}. Thus, our work paves the way for the quantum simulation of a broad range of phenomena, such as deconfined quantum criticality, entanglement in topologically ordered systems, and critical behaviors in open quantum systems \cite{vidal2003entanglement,senthil2004deconfined,ashida2017parity,xiao2019observation,liu2021non,shen2023proposal,rey2025incommensurate,yang2025gapless,yu2025gapless,yu2024universal,yang2025deconfined}.

\section*{ACKNOWLEDGMENTS}\label{sec}
We acknowledge the use of IBM Quantum services for this work. The views expressed are those of the authors and do not reflect the official policy or position of IBM or the IBM Quantum team. We acknowledge the helpful discussions with Prof. Ruben Verresen and Prof. Xuejia Yu. All data and codes of this work are available from the corresponding authors upon reasonable request. This work is supported by the Singapore Ministry of Education Academic Research Fund Tier-II Grant (Award No. MOE-T$2$EP$50222$-$0003$) and Tier-I preparatory grant (WBS no. A-8002656-00-00).

{\bf Author contributions.} R.~S. proposed the initial idea for this work and developed quantum algorithms. R.~S. and T.~C.  implemented simulations on the IBM Quantum devices. Z.~Y. contributed to the \ {classical} simulations, and Y.~B. provided theoretical analysis. C.~H.~L. is the overall supervisor for this project. 
All authors contributed to the preparation of the manuscript. 

{\bf Data availability.}
All data from this work are available from the corresponding authors upon reasonable request.

{\bf Code availability.}
All codes used in this work are available from the corresponding authors upon request. 

\section*{Methods}

\begin{figure}
	\centering
		\includegraphics[width=0.99\linewidth]{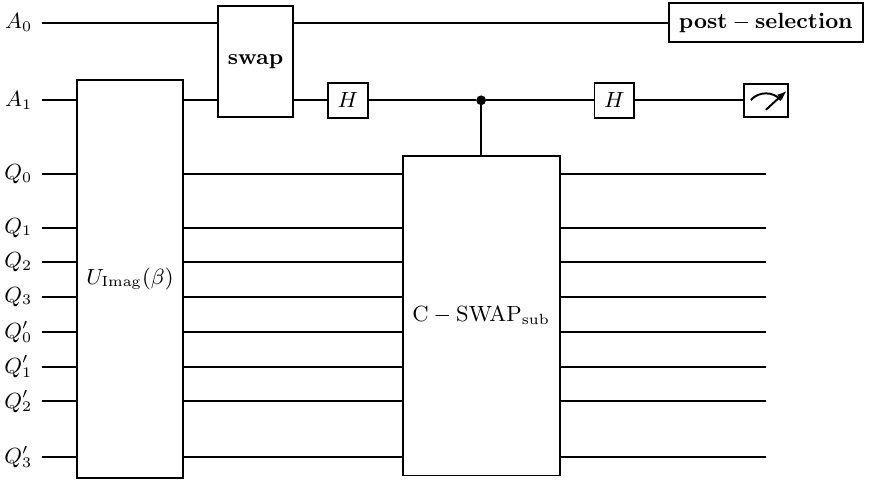}
	\caption{Quantum circuit for implementing the quantum amplitude estimation (QAE) method. $U_{\rm Imag}(\beta)$ is utilized for the ancilla-based preparation of ground states through the quantum imaginary-time evolution (QITE) method, specifically generating a 2-copy ground state via the operation $e^{-\beta H}\otimes e^{-\beta H}$. A swap gate transfers the post-selection from ancilla qubit $A_{1}$ to $A_{0}$. Subsequently, a controlled-swap operation, described by Eq.~\ref{cswap}, is applied to the selected subsystems $Q_0-Q_{x-1}$ and $Q^{\prime}_0-Q^{\prime}_{x-1}$ where $x$ represents the size of the selected subsystem. The second Renyi entropy is then measured as $2P(0)-1$ with $P(0)$ representing the probability of measuring $``0"$ at the $A_{1}$ ancilla qubit.
	}
	\label{fig:circuits}
\end{figure}

\subsection*{Circuits for quantum imaginary-time evolution}\label{QITE}
Here, we discuss our approach in simulating nonunitary processes $U(\beta)=e^{-\beta H}$, which represents imaginary time evolution. This nonunitary operation can be embedded into the following unitary operation $U_{\rm imag}(\beta)$ \cite{chen2022high,shen2311observation,lin2021real}:
\begin{equation}\label{nu}
	U_{\rm imag}(\beta)=\left[\begin{array}{cc}
		uU(\beta) & B \\                                                        
		C & D
	\end{array}\right].
\end{equation}
where $u^{-2}$ denotes the maximum eigenvalue of $U(\beta)$
and $C=A\sqrt{I-u^{2}\Sigma^{2}}B^{\dagger}$ with $I$ being the identity matrix. The matrices $A$, $B$, and $\Sigma$ are determined via singular value decomposition (SVD): $U(\beta)=A\Sigma B^{\dagger}$ ~\cite{shen2023observation,lin2021real}. The unitary $U_{\rm imag}(\beta)$ can be solved for by performing the QR decomposition of $U^{\prime\prime}$, described as below:
\begin{equation}\label{supussh}
	U_{\rm imag}(\beta)M=U^{\prime\prime},
\end{equation}
and
\begin{equation}
U^{\prime\prime}=\left[\begin{array}{cc}
		uU(\beta) & I\\           
		C& I
	\end{array}\right],
\end{equation}
where $M$ is an upper triangular matrix. For the results presented in the main text, we implement the QITE process using a trained variational circuit. Subsequent post-selection of the ancilla qubit representing the enlarged operator space recovers $U(\beta)$.

\subsection*{Quantum amplitude estimation}\label{qae}

We discuss a robust method for measuring the second Rényi entropy $R_{x}^{2}$, as described by Eq.~\ref{r2} of the main text. This measurement involves two steps: first, the preparation of a two-copy state, and second, the measurement of the second Rényi entropy using quantum amplitude estimation (QAE).

Here, we consider a $4$-qubit case for illustration [Fig.~\ref{fig:circuits}]. To prepare the corresponding 2-copy ground state, consisting of $8$ physical qubits, we can use the dilated operation $e^{-\beta H}\otimes e^{-\beta H}$, with each operation $e^{-\beta H}$ acting on a $4$-qubit chain, represented by $[Q_0, Q_1, Q_2, Q_3]$ and $[Q^{\prime}_0, Q^{\prime}_1, Q^{\prime}_2, Q^{\prime}_3]$. This QITE evolution can be implemented by coupling the system $[Q_0, Q_1, Q_2, Q_3, Q^{\prime}_0, Q^{\prime}_1, Q^{\prime}_2, Q^{\prime}_3]$ to an ancilla qubit.

After state preparation, a swap gate transfers the post-selection task from ancilla qubit $A_{1}$ to $A_{0}$  (see Fig.~\ref{fig:circuits}). This rearrangement ensures that QAE targets ancilla qubit $A_{1}$, optimally positioned near the physical chain.

Here, we use A to denote the selected subsystem, with its corresponding density matrix $\rho_{A}$. It can be shown that ${\rm Tr}(\rho_{A}^{2})={\rm Tr}({\rm SWAP}\rho_{A}^{\otimes 2})$ where the SWAP operator acts as ${\rm SWAP}\ket{\psi^{1}_{A}}\ket{\psi^{2}_{A}}=\ket{\psi^{2}_{A}}\ket{\psi^{1}_{A}}$ \cite{ekert2002direct}. Thus, the key quantity to measure is the expectation value of the SWAP gate acting on subsystem A. This can be realized through the QAE method. To measure an operator $O$ over $\ket{\psi}$, we first prepare a superposition state: $\frac{1}{\sqrt{2}}(\ket{\psi}\ket{0}+\ket{\psi}\ket{1})$. Applying a controlled-$O$ gate (controlled by the first qubit) followed by a Hadamard (H) gate acting on the first qubit results in the following state $\frac{1}{2}((I+O)\ket{\psi}\ket{0}+(I-O)\ket{\psi}\ket{1})$. Performing a $Z$-measurement on the first qubit leads to the probability of ``0": $P(0)=(1+\langle O \rangle)/2$, with $\langle O \rangle=2P(0)-1$.

We then utilize the following controlled-SWAP (C-SWAP) gate to implement the QAE process [FIG.~\ref{fig:circuits}]:
\begin{equation}\label{cswap}
	\text{\bf C-}[\operatorname{SWAP_{sub}(x)}]=I \otimes|0\rangle\langle 0|+	[\operatorname{SWAP_{sub}(x)}]\otimes| 1\rangle\langle 1|,
\end{equation}
where the operation $\operatorname{SWAP_{sub}(x)}$ represents the operation of swapping the selected length-$x$ subsystems, $Q_0-Q_{x-1}$ and $Q^{\prime}_0-Q^{\prime}_{x-1}$.

To obtain the final result, we only need to perform the post-selection on the $A_{0}$ qubit and simultaneously measure the $A_{1}$ qubit. Here, we can get the outcome $2P(0)-1$ where $P(0)$ denotes the probability of measuring $``0"$ at the $A_{1}$ qubit after post-selection. Such simulations are also realized on a trained ansatz circuit, and a representative circuit is shown in the following section.

\subsection*{Variational framework for robust simulations}\label{vqa}
In this section, we discuss the method for achieving robust simulations using our designed variational framework. This variational circuit serves a dual purpose: it significantly reduces circuit depth while providing a structured and adaptable framework for our enhanced error mitigation technique.

\begin{figure}
	\centering
\includegraphics[width=0.7\linewidth]{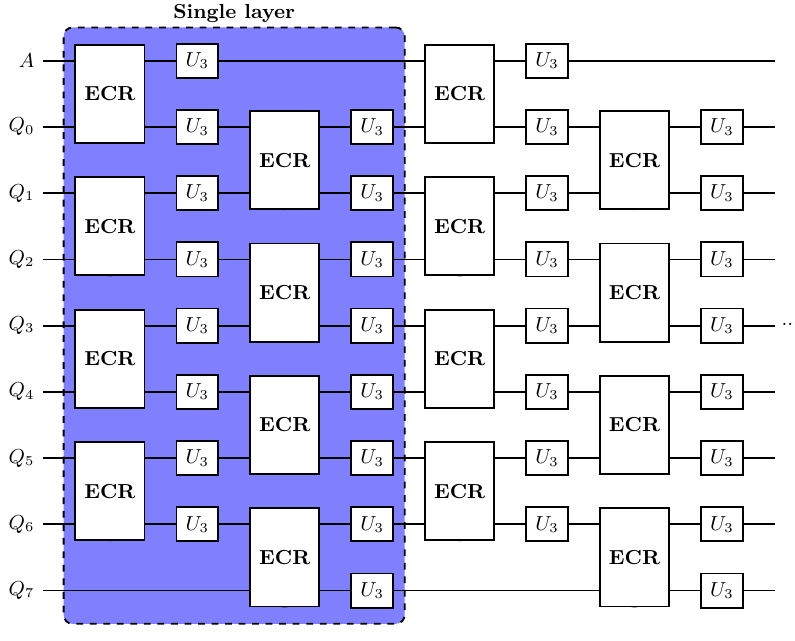}
	\caption{The structure of the variational circuit used. The blue block represents a single layer consisting of trainable $U_{3}$ gates and echoed cross-resonance (ECR) gates.  Each circuit layer of a 9-qubit chain is built by $8$ ECR gates, with an average error rate of $0.5\%$ (refer to Supplementary Information for details).}
	\label{circuit2}
\end{figure}

We first introduce our variational circuit $V$ and present a representative $8$-qubit case in FIG.~\ref{circuit2}. The circuit is constructed using two fundamental elements. The echoed cross-resonance (ECR) gate is utilized as the primary entangling gate, defined by the following matrix:
\begin{equation}
{\rm E C R} =\frac{1}{\sqrt{2}}\left(\begin{array}{cccc}
0 & 1 & 0 & i \\
1 & 0 & -i & 0 \\
0 & i & 0 & 1 \\
-i & 0 & 1 & 0
\end{array}\right).
\end{equation}
Each variational layer includes trainable $U_{3}$ gate\CH{s}, each defined as:
\begin{equation}
	U_{3}(\theta, \phi, \lambda)=\left[\begin{array}{cc}
		\cos \left(\frac{\theta}{2}\right) & -e^{i \lambda} \sin \left(\frac{\theta}{2}\right) \\
		e^{i \phi} \sin \left(\frac{\theta}{2}\right) & e^{i(\phi+\lambda)} \cos \left(\frac{\theta}{2}\right)
	\end{array}\right].
\end{equation}
The parameters $(\theta, \phi, \lambda)$ are optimized during the training process to approach the target state.

To simulate a process $U$ on circuits, we minimize the following cost function to zero:
\begin{equation}\label{E}
	C(\theta, \phi, \lambda)=1-F(V_{n}(\theta, \phi, \lambda)\ket{\psi}, U\ket{\psi}),
\end{equation}
where $U$ represents the original circuit used in our work, $\ket{\psi}$ is the required initial state, and $V_{n}$ is the  variational circuit consisting of $n$ layers (see FIG.~\ref{circuit2}). The function $F(\ket{a},\ket{b})=|\langle a|b \rangle|$ is the measurement of the overlap between two input states. For the preparation of ground states, we set $U=U_{\rm imag}(\beta)$, leading to a modified cost function:
\begin{equation}
	C(\theta, \phi, \lambda)=1-F(V_{n}(\theta, \phi, \lambda)\ket{\psi},{\rm Post}[U\ket{\psi}]).
\end{equation}
Here, ``${\rm Post}$" denotes the post-selection operation, which effectively mitigates measurement loss. To further implement the process of quantum amplitude estimation, we employ a similar cost function:
\begin{equation}
	C(\theta, \phi, \lambda)=1-F(V_{n}(\theta, \phi, \lambda)\ket{\psi},U_{\rm QAE}{\rm Post}[U\ket{\psi}]),
\end{equation}
where $U_{\rm QAE}$ denotes the QAE circuit structure, as shown in FIG.~\ref{fig:circuits}.

For simulations in our work based on such variational circuits, each data point is obtained by averaging the results from $20000$ runs on the IBM Quantum device.

Another critical step in our simulation is the implementation of the zero noise extrapolation (ZNE) procedure. To integrate ZNE into our variational circuits, we extend the original trained  circuit $V_{n}$ as follows, as demonstrated in FIG.~\ref{circuit} (b),
\begin{equation}\label{E}
V^{\rm ZNE}_{n+m}=V^{\rm Id}_{m}V_{n}.
\end{equation}
Here, $V^{\rm Id}_{m}$ represents an identity circuit composed of $m$ additional layers that follow the structure of original variational layers without changing the computational outcome in the noiseless limit. However, due to inevitable device noise, increasing the number of layers also provides a tuning knob in controlling the extent of noise, whose effect can be extrapolated down to zero. Consequently, our improved variational circuit design significantly enhances noise management and mitigation, allowing for more precise measurements of topological properties even under the constraint of current noisy quantum hardware.

\subsection{Quantum state tomography}
Here, we introduce our implementation of quantum state tomography \cite{cramer2010efficient,christandl2012reliable}. In the main text, we consider measuring the reduced density matrix of three qubits in a length-4 chain. The corresponding reduced density matrix is represented as a sum of Pauli string operators:
\begin{equation}\label{rho}
	\rho_{ {A}}=\frac{1}{8}\sum_{S_{1},S_{2},S_{3}=I,X,Y,Z }c_{1,2,3}S_{1}S_{2}S_{3}.
\end{equation}
These coefficients $c_{1,2,3}$ are measured on a quantum computer along the basis $S_{1}S_{2}S_{3}$:
\begin{equation}
	c_{1,2,3}=\operatorname{Tr}[S_{1}S_{2}S_{3}\rho_{ {A}}].
\end{equation}

Since the identity $I$ is trivial in any Pauli string, the total $4^{3}=64$ coefficients can be determined through $3^{3}=27$ measurements of the first three physical qubits along $X, Y, Z$ directions.  For noisy simulations on the IBM Quantum device, we first collect coefficients $c_{M, N}$ obtained from the raw data of $27$ executions and then generate the corresponding $64$ Pauil strings to construct this reduced density matrix. Then, we compute the eigenstates of this reduced density matrix.

\subsection{Symmetry-protected edge states and entanglement entropy}
To illustrate the robustness of the topological entanglement entropy and edge states in the cluster SPT phase, we examine the cluster model   $H_{c}=\sum_iZ_{i-1} X_i Z_{i+1}$ in a minimal 4-qubit system. Its ground state can be expressed as:
\begin{equation}
    \ket{\psi_{L=4}}=\frac{1}{2}(\ket{0000}+\ket{0100}+\ket{0010}-\ket{0110}).
\end{equation}
In this state, the edge spins are perfectly isolated. The half-chain reduced density matrix of this system yields two identical eigenvalues, $\epsilon_{1}=\epsilon_{2}=1/2$, resulting in the entanglement entropy value of 
 $\ln(2)$. Remarkably, this characteristic entropy emerges even in such a small system.
Next, we show how this result remains largely robust in the presence of significant Ising interaction perturbations. Under such perturbations, we can still approximate the ground state as:
\begin{align}
        \ket{\psi_{L=4}}^{\prime}=\frac{1}{2}((1+\delta^{\prime}_{1})\ket{0000}+(1-\delta^{\prime}_{2})\ket{0100}\\
    +(1-\delta^{\prime}_{3})\ket{0010}-(1-\delta^{\prime}_{4})\ket{0110}).
\end{align}
where $\delta^{\prime}_{i}$ are small corrections induced by the Ising interaction. Crucially, the corresponding eigenvalues of this density matrix can be represented as: $\epsilon_{1}\approx1/2+\delta$, $\epsilon_{2}\approx1/2-\delta$, and $\epsilon_{3,4}\approx 0$, which still exhibits a salient entanglement gap. Here, $\delta$ scales with $\delta^{\prime}_{i}$. Therefore, despite such perturbations, the fundamental structure of edge states remains stable, and the topological entanglement remains robust. 

\bibliography{ref}

\onecolumngrid
\flushbottom
\newpage
\appendix
\setcounter{equation}{0}
\setcounter{figure}{0}
\setcounter{table}{0}
\setcounter{section}{0}
\renewcommand{\theequation}{S\arabic{equation}}
\renewcommand{\thefigure}{S\arabic{figure}}
\renewcommand{\thesection}{S\arabic{section}}
\renewcommand{\thepage}{S\arabic{page}}

\newpage
\section*{Supplementary Informations}
\section{IBM Q Quantum hardware}\label{ibmq}
We present data from the ``ibm-brisbane" device, which has a total of $127$ qubits. For simulations in this work, we utilize the qubit chain $[10,9,8,7,6,5,4,3,2,1,0]$ chosen for its optimal noise levels. The readout error rate of ancilla qubits 9 and 10 is approximately $0.6\%$. This low error rate ensures robust simulations, which require ancilla post-selection.
\begin{figure}[h]
	\centering
	\includegraphics[width=0.7\linewidth]{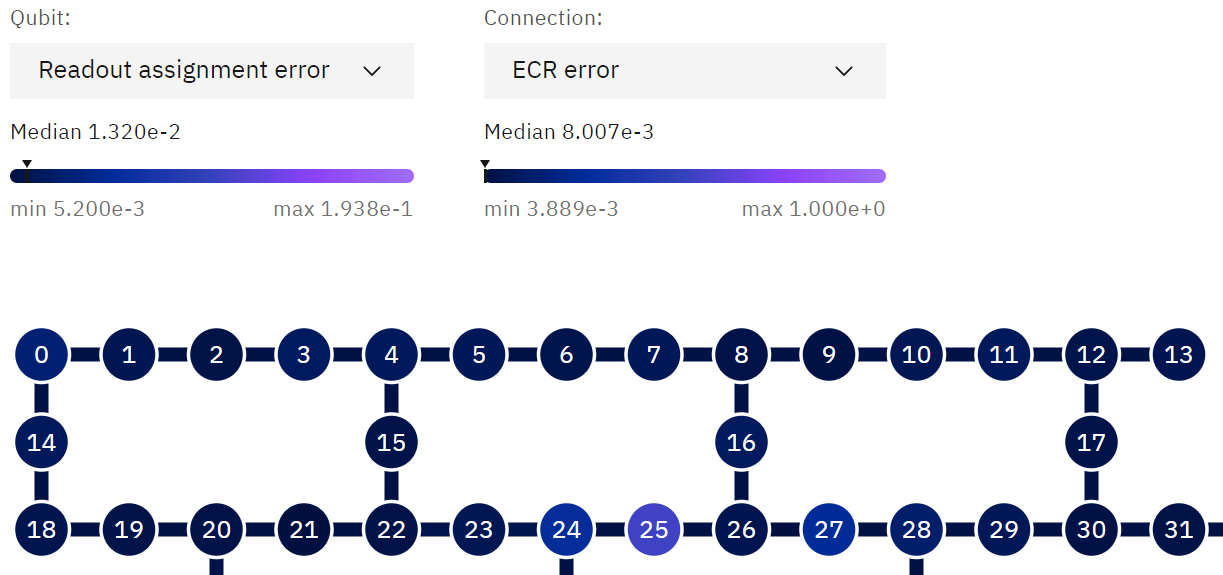}
	\caption{Noise data of the ''ibm-brisbane" device. The order of the selected qubit chain is $[10,9,8,7,6,5,4,3,2,1,0]$. The average ECR gate error rate for the selected connections is approximately $0.5\%$. Additionally, for the ancilla qubits $9$ and $10$, the readout error rate is $0.6\%$.}
	\label{fig:ibmq2}
\end{figure}

\subsection{Error mitigation}
In our work, we employ the method of zero-noise extrapolation (ZNE) \cite{he2020zero} to mitigate noise effects. This technique involves amplifying noise levels in quantum simulations and then using these data points to extrapolate back to a theoretical zero-noise limit.

\begin{figure}[h]
	\centering
	\begin{tikzpicture}
		\node[scale=0.6] {
			\begin{quantikz}
				&\lstick{$A$}~~&\gate[2]{\bf ECR}\gategroup[9,steps=4,style={dashed,rounded corners,fill=blue!50, inner xsep=2pt},background]{{\bf Original trained layers}}&\gate{U_{3}}&\qw&\qw &\gate[2]{\bf ECR}\gategroup[9,steps=10,style={dashed,rounded corners,fill=green!50, inner xsep=2pt},background]{{\bf Identity block with additional layers}}&\gate{U_{3}}&\qw&\qw&\qw&\gate[2]{\bf ECR}&\gate{U_{3}}&\qw&\qw&\qw\text{......}\\
				&\lstick{$Q_{0}$}&\targ{ }&\gate{U_{3}}&\gate[2]{\bf ECR}&\gate{U_{3}}&\targ{ }&\gate{U_{3}}&\gate[2]{\bf ECR}&\gate{U_{3}}&\qw&\targ{ }&\gate{U_{3}}&\gate[2]{\bf ECR}&\gate{U_{3}}&\qw\text{......}\\
				&\lstick{$Q_{1}$}&\gate[2]{\bf ECR}&\gate{U_{3}}&\targ{ }&\gate{U_{3}}&\gate[2]{\bf ECR}&\gate{U_{3}}&\targ{ }&\gate{U_{3}}&\qw&\gate[2]{\bf ECR}&\gate{U_{3}}&\targ{ }&\gate{U_{3}}&\qw\text{......}\\
				&\lstick{$Q_{2}$}&\targ{ }&\gate{U_{3}}&\gate[2]{\bf ECR}&\gate{U_{3}}&\targ{ }&\gate{U_{3}}&\gate[2]{\bf ECR}&\gate{U_{3}}&\qw&\targ{ }&\gate{U_{3}}&\gate[2]{\bf ECR}&\gate{U_{3}}&\qw\text{......}\\
				&\lstick{$Q_{3}$}&\gate[2]{\bf ECR}&\gate{U_{3}}&\targ{ }&\gate{U_{3}}&\gate[2]{\bf ECR}&\gate{U_{3}}&\targ{ }&\gate{U_{3}}&\qw&\gate[2]{\bf ECR}&\gate{U_{3}}&\targ{ }&\gate{U_{3}}&\qw\text{......}\\
				&\lstick{$Q_{4}$}&\targ{ }&\gate{U_{3}}&\gate[2]{\bf ECR}&\gate{U_{3}}&\targ{ }&\gate{U_{3}}&\gate[2]{\bf ECR}&\gate{U_{3}}&\qw&\targ{ }&\gate{U_{3}}&\gate[2]{\bf ECR}&\gate{U_{3}}&\qw\text{......}\\
				&\lstick{$Q_{5}$}&\gate[2]{\bf ECR}&\gate{U_{3}}&\targ{ }&\gate{U_{3}}&\gate[2]{\bf ECR}&\gate{U_{3}}&\targ{ }&\gate{U_{3}}&\qw&\gate[2]{\bf ECR}&\gate{U_{3}}&\targ{ }&\gate{U_{3}}&\qw\text{......}\\
				&\lstick{$Q_{6}$}&\targ{ }&\gate{U_{3}}&\gate[2]{\bf ECR}&\gate{U_{3}}&\targ{ }&\gate{U_{3}}&\gate[2]{\bf ECR}&\gate{U_{3}}&\qw&\targ{ }&\gate{U_{3}}&\gate[2]{\bf ECR}&\gate{U_{3}}&\qw\text{......}\\
				&\lstick{$Q_{7}$}&\qw&\qw&\targ{}&\gate{U_{3}}&\qw&\qw&\targ{ }&\gate{U_{3}}&\qw&\qw&\qw&\targ{ }&\gate{U_{3}}&\qw\text{......}
			\end{quantikz}
		};
	\end{tikzpicture}
	\caption{Circuit for implementing the method of zero-noise extrapolation (ZNE). The blue block represents the trained circuit used for conducting physical simulations of the quantum imaginary-time evolution or other processes. The green block, consisting of additional layers, serves as an identity function to amplify the impact of gate errors.}
	\label{circuit3}
\end{figure}
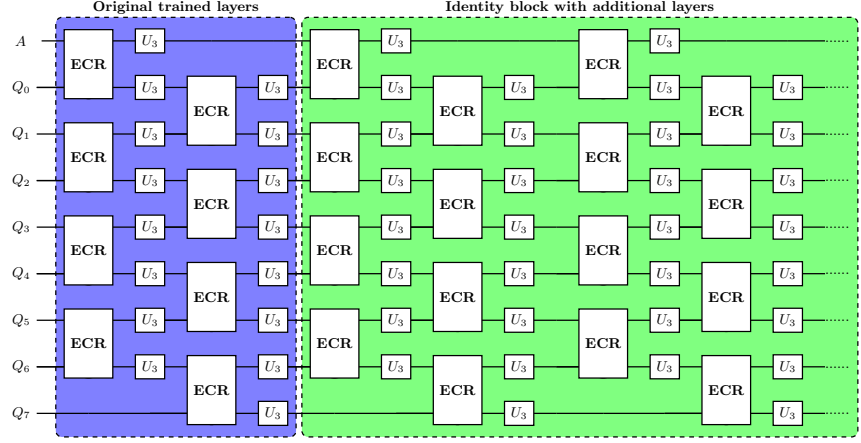

We introduce a more scalable framework based on variational circuits, as illustrated in FIG.\ref{circuit3}. The initial blue block represents the original trained circuit. Subsequently, additional green identity blocks are introduced to amplify noise effects and simulate conditions for increased error.

\begin{figure}[h]
	\centering
	\includegraphics[width=0.9\linewidth]{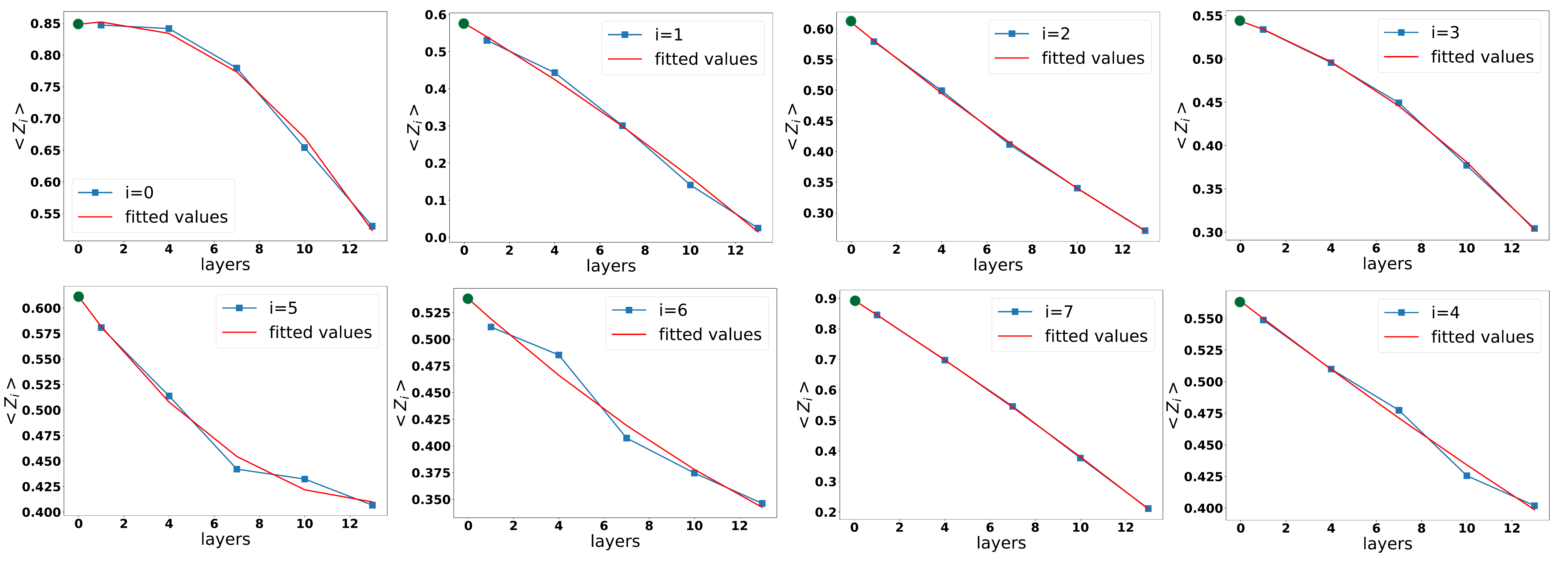}
	\caption{Illustration of the implementation of the zero-noise extrapolation method. These data on the magnetizations on sites labeled by different positions $i$ correspond to the results shown in FIG. 4 of the main text. The entire circuit consists of trained layers combined with identity layers.  The zero-noise results (green dots) are obtained by extrapolating the noisy data, represented by the blue dots, to the zero-layer point using the fitted red curves. 
    }
	\label{fig:mitresult}
\end{figure}

In FIG.~\ref{fig:mitresult}, we present a representative case that demonstrates the implementation of the ZNE method to achieve the results shown in FIG. 4 of the main text. This process involves performing noisy simulations across different circuit layers, including both original and identity layers. Since the structures of all layers are identical, the number of layers serves as an appropriate noise amplification factor. These results of different layers are indicated by the blue dots in FIG.~\ref{fig:mitresult}). Following this, the zero-noise results are extrapolated from these noisy data points by fitting them to the red curves and extending the fit to the zero-layer limit. These results illustrate how the ZNE method mitigates noise, enabling our simulations to approach theoretical zero-noise conditions.

\subsection{Sampling errors in simulators}
\begin{figure}[h]
\centering
\includegraphics[width=0.7\linewidth]{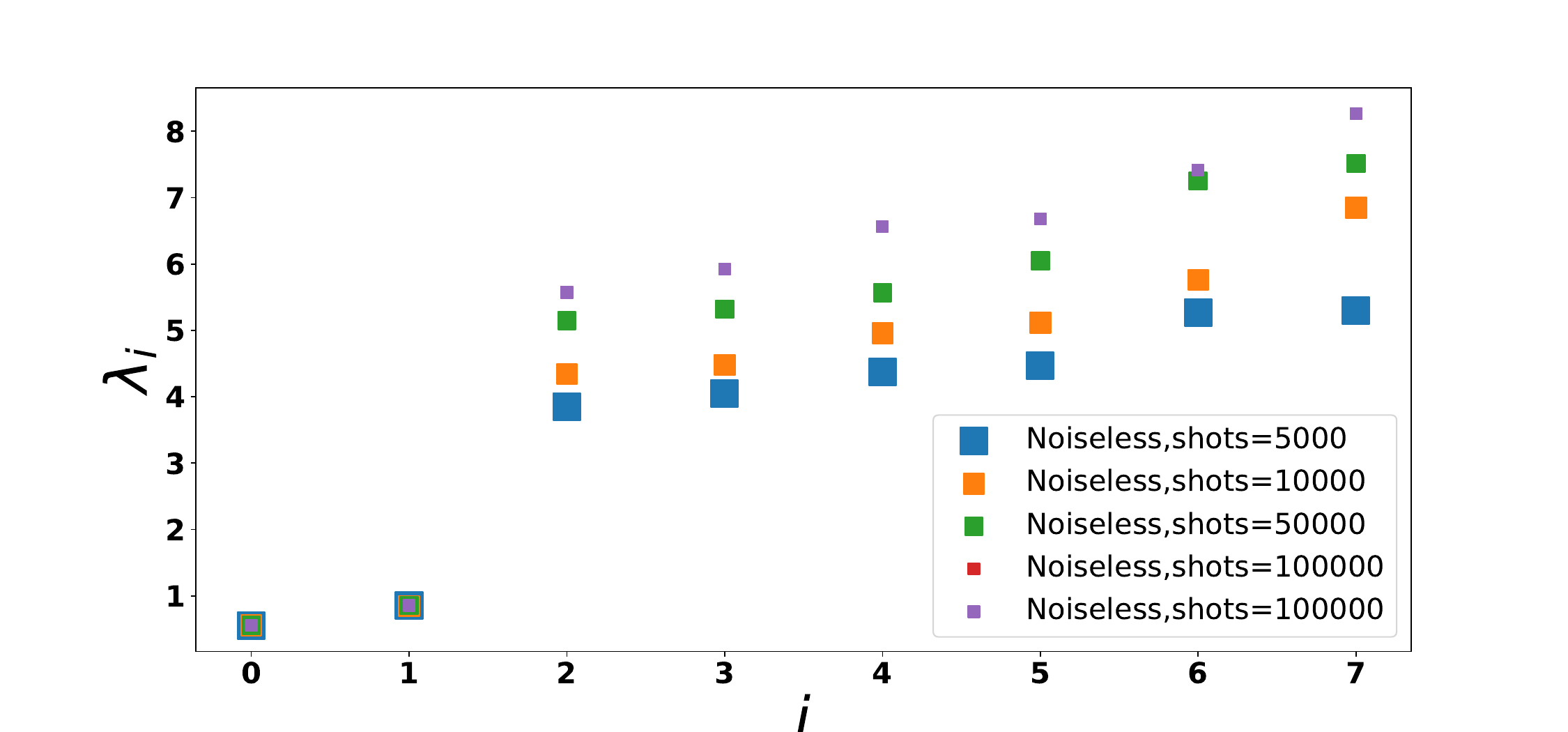}
\caption{Estimation of the sampling error in quantum state tomography. We classically measure the entanglement spectrum of the 3-qubit subspace within a 4-qubit system using noiseless circuits. The reduced density matrix is constructed using the method described by Eq.\ref{rho} (main text). The squares represent results simulated on the Qiskit noiseless simulator, which includes sampling noise. Notably, increasing the number of measurement shots does not lead to significant improvement. Other parameters are $g=2.5$ and $J=h=1$.}
\label{fig:sampleerror}
\end{figure}
As discussed in the main text, we demonstrate that significant sampling errors emerge in the measurement of a density matrix on quantum computers. To understand these errors further, we examine the impact of the number of measurement shots, as shown in FIG.~\ref{fig:sampleerror}. The entanglement gap becomes evident as the number of shots increases. However, a substantial increase in the number of shots from $5000$ to $200000$ does not yield significant improvements, highlighting the inevitability of sampling errors on current quantum processors.

\flushbottom

\end{document}